\documentclass[preprint,aps]{revtex4} 

\usepackage{epsf}

\begin{document}

\title{High-Order Coupled Cluster Method  Calculations for the Ground- 
and Excited-State Properties of the Spin-Half {\it XXZ} Model}

\author{R. F. Bishop,$^a$ D. J. J. Farnell,$^a$ S.E. Kr\"uger,$^b$
J. B. Parkinson,$^a$ J. Richter,$^b$ and C. Zeng.$^{a,}$\footnote{Presently at:
Department of Physics and Astronomy, Rutgers University, Piscataway, 
NJ 08855, USA.}}      
\affiliation{$^a$Department of Physics, University of Manchester Institute of
  Science and Technology (UMIST), P O Box 88, Manchester M60 1QD, United 
  Kingdom}
\affiliation{$^b$Institut f\"ur Theoretische Physik, Universit\"at Magdeburg, 
  P.O. Box 4120, D-39016 Magdeburg, Germany}

\date{\today}

\begin{abstract}
In this article, we present new results of high-order coupled cluster 
method (CCM) calculations, based on a N\'eel model state with spins 
aligned in the $z$-direction, for both the ground- and excited-state 
properties of the spin-half {\it XXZ} model on the linear chain, the 
square lattice, and the simple cubic lattice. 
In particular, the high-order CCM formalism is extended to treat 
the excited states of lattice quantum spin systems for the first time. 
Completely new results for the excitation energy gap of the spin-half 
{\it XXZ} model for these lattices are thus determined. 
These high-order calculations are based on a localised approximation 
scheme called the LSUB$m$ scheme in which we retain all $k$-body 
correlations defined on all possible locales of $m$ adjacent 
lattice sites ($k \le m$). 
The ``raw'' CCM LSUB$m$ results are seen to provide very good results 
for the ground-state energy, sublattice magnetisation, and the value 
of the lowest-lying excitation energy for each of these systems. 
However, in order to obtain even better results, two types of 
extrapolation scheme of the LSUB$m$ results to the limit $m 
\rightarrow \infty$ (i.e., the exact solution in the thermodynamic 
limit) are presented.
The extrapolated results provide extremely accurate results for the 
ground- and excited-state properties of these systems across a wide 
range of values of the anisotropy parameter.

\begin{flushleft}
\end{flushleft}
\end{abstract}
\maketitle

\section{Introduction}

The coupled cluster method (CCM) 
 \cite{newccm1,newccm2,newccm3,newccm4,newccm5,newccm6,newccm7,newccm8} 
has previously been applied  \cite{ccm1,ccm2,ccm3,ccm4,ccm5,ccm6,ccm7,ccm8,ccm9,ccm10,ccm11,ccm12,ccm13,ccm14} 
to various lattice quantum spin systems with much success. 
In particular, high-order  \cite{ccm7,ccm8,ccm9,ccm11,ccm12,ccm13}
CCM calculations using a localised approximation scheme
called the LSUB$m$ approximation scheme have previously been 
utilised to determine the ground-state properties of the 
spin-half square lattice {\it XXZ} model to high accuracy 
(see especially Refs.  \cite{ccm7,ccm11}). These calculations 
were performed both in a regime characterised by non-zero N\'eel 
order \cite{ccm1,ccm2,ccm3,ccm7} in the $z$-direction, and also in a 
regime characterised by non-zero N\'eel ordering in the $xy$-plane
 \cite{ccm8,ccm11}. The results were found to be in excellent 
agreement with the known results of the {\it XXZ} model
(discussed below). 

The spin-half {\it XXZ} model is defined by the following Hamiltonian,
\begin{equation}
H = \frac 12 \sum_{i=1}^N \sum_{\rho}\left[ s_i^x s_{i+\rho}^x  + 
s_i^y s_{i+\rho}^y  + \Delta s_i^z s_{i+\rho}^z \right] ~~ ,
\label{XXZ_hamiltonian}
\end{equation}  
where the index {\it i} in Eq. (\ref{XXZ_hamiltonian}) 
runs over all $N$ lattice 
sites, and $\rho$ runs over all {\it z} nearest-neighbours 
lattice vectors with respect to $i$ on the chain ({\it z}=2), 
the square lattice ({\it z}=4), and the simple cubic lattice 
({\it z}=6). Periodic boundary conditions are also 
assumed. We note that $[H,s_T^z]=0$ for this model, where
$s_T^z = \sum_{i=1}^N s_i^z$. 
The spin-half {\it XXZ} model contains three regimes with respect 
to the anisotropy parameter $\Delta$. 
For $\Delta \le -1$ there is a ferromagnetic 
regime in which the classical ground state is also 
the quantum-mechanical ground state. At $\Delta=-1$, there is a
first-order phase transition to a planar regime in which
the classical ground state is a N\'eel state in the $xy$-plane. 
For the square and cubic lattices, the quantum-mechanical 
ground state in this regime is furthermore characterised 
by non-zero, in-plane, long-range ordering. 
For the linear chain quantum fluctuations are strong enough to 
destroy all long-range ordering and so this region in 1D
is sometimes referred to as the ``critical regime.''
In the 1D case the planar regime spans the range $-1<\Delta<1$, 
and at $\Delta=1$ there is another phase transition 
to a phase characterised by non-zero N\'eel ordering 
in the $z$-direction for the ground state. 
For the linear chain, the Bethe Ansatz \cite{ba1,ba2,ba3,ba4} 
provides an exact solution of the quantum spin-half {\it XXZ} model,
although for higher spatial dimensionality no 
exact solutions have as yet been determined. However, extensive 
approximate calculations have been carried out for this 
model, in particular for the Heisenberg antiferromagnet 
(HAF) at $\Delta=1$. 
It is believed that the 1D phase transition at $\Delta=1$
also occurs for higher spatial dimensionality than one. 
Examples of such approximate calculations 
for the spin-half {\it XXZ} and HAF models are spin-wave 
theory \cite{swt1,swt2} (SWT), exact diagonalisations of 
finite-sized systems \cite{finite1,finite2}, exact cumulant series 
expansions \cite{series1}, and quantum Monte 
Carlo \cite{qmc1,qmc2,qmc3,qmc4} (QMC) calculations. In particular,
we note that such results for the spin-half isotropic HAF 
on the square lattice typically predict that approximately 
61-62$\%$ of the classical N\'eel-ordering remains in the 
quantum case.

In this article we present results of high-order coupled 
CCM calculations for the ground- and excited-state properties of the 
spin-half {\it XXZ} model on the linear chain, square lattice, and 
simple cubic lattice based on the systematic incorporation of many-spin
correlations on top of a N\'eel model state with spins aligned in the 
$z$-direction. In Sec. II, we describe the high-order CCM 
formalisms for both the ground and excited states. 
Although the high-order CCM formalism has been described 
previously  \cite{ccm11}, we present a very brief overview 
of it in order to give a necessary background for the new high-order 
formalism for the excited state. We also discuss in Sec. 
II the manner in which the CCM results for a localised approximation 
scheme called the LSUB$m$ scheme (in which we retain all $k$-body 
correlations defined on all possible locales of $m$ adjacent lattice 
sites ($k \le m$)) are extrapolated to the limit $m \rightarrow 
\infty$. In Sec. III, we present the results of our high-order
treatment of the {\it XXZ} model for these lattices, and finally
our conclusions are presented in Sec. IV. 

\section{The Coupled Cluster Method (CCM)}

In this section, we firstly describe the general ground-state CCM formalism 
\cite{newccm1,newccm2,newccm3,newccm4,newccm5,newccm6,newccm7,newccm8},
and then show how to apply it to the specific case of the spin-half 
{\it XXZ} model. This is then extended to deal with excited states.

\subsection{The Ground-State Formalism}

The exact ket and bra ground-state energy 
eigenvectors, $|\Psi\rangle$ and $\langle\tilde{\Psi}|$, of a 
many-body system described by a Hamiltonian $H$, 
\begin{equation} 
H |\Psi\rangle = E_g |\Psi\rangle
\;; 
\;\;\;  
\langle\tilde{\Psi}| H = E_g \langle\tilde{\Psi}| 
\;, 
\label{eq1} 
\end{equation} 
are parametrised within the single-reference CCM as follows:   
\begin{eqnarray} 
|\Psi\rangle = {\rm e}^S |\Phi\rangle \; &;&  
\;\;\; S=\sum_{I \neq 0} {\cal S}_I C_I^{+}  \nonumber \; , \\ 
\langle\tilde{\Psi}| = \langle\Phi| \tilde{S} {\rm e}^{-S} \; &;& 
\;\;\; \tilde{S} =1 + \sum_{I \neq 0} \tilde{{\cal S}}_I C_I^{-} \; .  
\label{eq2} 
\end{eqnarray} 
The single model or reference state $|\Phi\rangle$ is required to have the 
property of being a cyclic vector with respect to two well-defined Abelian 
subalgebras of {\it multi-configurational} creation operators $\{C_I^{+}\}$ 
and their Hermitian-adjoint destruction counterparts $\{ C_I^{-} \equiv 
(C_I^{+})^\dagger \}$. Thus, $|\Phi\rangle$ plays the role of a vacuum 
state with respect to a suitable set of (mutually commuting) many-body 
creation operators $\{C_I^{+}\}$, 
\begin{equation} 
C_I^{-} |\Phi\rangle = 0 \;\; , \;\;\; I \neq 0 \; , 
\label{eq3}
\end{equation} 
with $C_0^{-} \equiv 1$, the identity operator. These operators are 
complete in the many-body Hilbert (or Fock) space,  
\begin{equation} 
1=|\Phi\rangle \langle\Phi| + \sum_{I\neq 0} 
C_I^{+}  |\Phi\rangle \langle\Phi| C_I^{-} \; . 
\label{eq4}
\end{equation} 
The {\it correlation operator} $S$ is decomposed entirely in terms 
of these creation operators $\{C_I^{+}\}$, which, when acting on the 
model state ($\{C_I^{+}|\Phi\rangle \}$), create multi-particle 
excitations on top of the model state. We note that although the 
manifest Hermiticity, ($\langle \tilde{\Psi}|^\dagger = 
|\Psi\rangle/\langle\Psi|\Psi\rangle$), is lost in these 
parametrisations, the intermediate normalisation condition 
$ \langle \tilde{\Psi} | \Psi\rangle
= \langle \Phi | \Psi\rangle 
= \langle \Phi | \Phi \rangle \equiv 1$ is explicitly 
imposed. The {\it correlation coefficients} $\{ {\cal S}_I, \tilde{{\cal S}}_I \}$ 
are regarded as being independent variables, even though formally 
we have the relation, 
\begin{equation} 
\langle \Phi| \tilde{S} =
\frac{ \langle\Phi| {\rm e}^{S^{\dagger}} {\rm e}^S } 
     { \langle\Phi| {\rm e}^{S^{\dagger}} {\rm e}^S |\Phi\rangle } \; . 
\label{eq5}
\end{equation} 
The full set $\{ {\cal S}_I, \tilde{{\cal S}}_I \}$ thus provides a complete 
description of the ground state. For instance, an arbitrary 
operator $A$ will have a ground-state expectation value given as, 
\begin{equation} 
\bar{A}
\equiv \langle\tilde{\Psi}\vert A \vert\Psi\rangle
=\langle\Phi | \tilde{S} {\rm e}^{-S} A {\rm e}^S | \Phi\rangle
=\bar{A}\left( \{ {\cal S}_I,\tilde{{\cal S}}_I \} \right) 
\; .
\label{eq6}
\end{equation} 

We note that the exponentiated form of the ground-state CCM 
parametrisation of Eq. (\ref{eq2}) ensures the correct counting of 
the {\it independent} and excited correlated 
many-body clusters with respect to $|\Phi\rangle$ which are present 
in the exact ground state $|\Psi\rangle$. It also ensures the 
exact incorporation of the Goldstone linked-cluster theorem, 
which itself guarantees the size-extensivity of all relevant 
extensive physical quantities. 

The determination of the correlation coefficients $\{ {\cal S}_I, \tilde{{\cal S}}_I \}$ 
is achieved by taking appropriate projections onto the ground-state 
Schr\"odinger equations of Eq. (\ref{eq1}). Equivalently, they may be 
determined variationally by requiring the ground-state energy expectation 
functional $\bar{H} ( \{ {\cal S}_I, \tilde{{\cal S}}_I\} )$, defined as in Eq. (\ref{eq6}), 
to be stationary with respect to variations in each of the (independent) 
variables of the full set. We thereby easily derive the following coupled 
set of equations, 
\begin{eqnarray} 
\delta{\bar{H}} / \delta{\tilde{{\cal S}}_I} =0 & \Rightarrow &   
\langle\Phi|C_I^{-} {\rm e}^{-S} H {\rm e}^S|\Phi\rangle = 0 ,  \;\; 
\forall I \neq 0 \;\; ; \label{eq7} \\ 
\delta{\bar{H}} / \delta{{\cal S}_I} =0 & \Rightarrow & 
\langle\Phi|\tilde{S} {\rm e}^{-S} [H,C_I^{+}] {\rm e}^S|\Phi\rangle 
= 0 , \;\; \forall I \neq 0 \;\; . \label{eq8}
\end{eqnarray}  
Equation (\ref{eq7}) also shows that the ground-state energy at the stationary 
point has the simple form 
\begin{equation} 
E_g = E_g ( \{{\cal S}_I\} ) = \langle\Phi| {\rm e}^{-S} H {\rm e}^S|\Phi\rangle
\;\; . 
\label{eq9}
\end{equation}  
It is important to realize that this (bi-)variational formulation 
does {\it not} lead to an upper bound for $E_g$ when the summations for 
$S$ and $\tilde{S}$ in Eq. (\ref{eq2}) are truncated, due to the lack of 
exact Hermiticity when such approximations are made. However, it is clear 
that the important Hellmann-Feynman theorem {\it is} preserved in all 
such approximations. 

We also note that Eq. (\ref{eq7}) represents a coupled set of 
nonlinear multinomial equations for the {\it c}-number correlation 
coefficients $\{ {\cal S}_I \}$. The nested commutator expansion 
of the similarity-transformed Hamiltonian,  
\begin{equation}  
\hat H \equiv {\rm e}^{-S} H {\rm e}^{S} = H 
+ [H,S] + {1\over2!} [[H,S],S] + \cdots 
\;\; , 
\label{eq10}
\end{equation} 
together with the fact that all of the individual components of 
$S$ in the sum in Eq. (\ref{eq2}) commute with one another, imply 
that each element of $S$ in Eq. (\ref{eq2}) is linked directly to
the Hamiltonian in each of the terms in Eq. (\ref{eq10}). Thus,
each of the coupled equations (\ref{eq7}) is of linked-cluster type.
Furthermore, each of these equations is of finite length when expanded, 
since the otherwise infinite series of Eq. (\ref{eq10}) will always 
terminate at a finite order, provided (as is usually the case) 
only that each term in the 
second-quantised form of the Hamiltonian $H$ contains a finite number of 
single-body destruction operators, defined with respect to the reference 
(vacuum) state $|\Phi\rangle$. Therefore, the CCM parametrisation naturally 
leads to a workable scheme which can be efficiently implemented 
computationally. It is also important to note that at the heart
of the CCM lies a similarity transformation, in contrast with  
the unitary transformation in a standard variational formulation 
in which the bra state $\langle\tilde\Psi|$ is simply taken as
the explicit Hermitian adjoint of $|\Psi\rangle$. 

We now wish to apply the general CCM formalism outlined
above to the specific case of the spin-half {\it XXZ} model, 
and we choose the N\'eel state, in which the spins lie 
along the $z$-axis, to be the model state. Furthermore, 
we perform a rotation of the local axes of the up-pointing 
spins by 180$^\circ $ about the $y$ axis, so
that spins on both sublattices may be treated
equivalently. The (canonical) transformation is described by,
\begin{equation}
s^x \; \rightarrow \; -s^x, \; s^y \; \rightarrow \;  s^y, \;
s^z \; \rightarrow \; -s^z  \; .
\end{equation}
The model state now appears $mathematically$ to consist
of purely down-pointing spins in these rotated local
axes. In terms of the spin raising and lowering operators
$s_k^{\pm} \equiv s_k^x \pm i s_k^y$ the Hamiltonian 
may be written in these local axes as,
\begin{equation}
H = -\frac 14 \sum_i^N \sum_{\rho} \; \biggl[ \; s_i^+
s_{i+\rho}^+ + s_i^-s_{i+\rho}^- + 2 \Delta s_i^z s_{i+\rho}^z  \; \biggr] \; .
\label{eq:newH}
\end{equation}
In this article, we also wish to perform high-order CCM calculations 
for this model, and in order to do this we firstly define the
$\{ C_I^+ \}$ operators more explicitly, as $C_I^+ \equiv 
s^+_{i_1} s^+_{i_2}  \cdot\cdot\cdot s^+_{i_l}$, where the 
set-index $I\equiv \{i_1, i_2, ... , i_l\}$. Each of the 
single-site indices is allowed to cover {\it all} lattice 
sites, although double (or greater) occupancy of any particular 
site in any set $I$ is explicitly prohibited 
for the spin-half case, since $(s^+)^2=0$. We analogously
define the ket-state correlation coefficients as 
${\cal S}_I \equiv {\cal S}_{i_1,i_2,\cdot\cdot\cdot,i_l}$ 
for a cluster $I$ defined above. By construction the 
coefficients ${\cal S}_{i_1,i_2,\cdot\cdot\cdot,i_l}$ 
are completely symmetric under the interchange of any
two indices. It is also useful to define the following 
two operators, 
\begin{eqnarray}
F_k  &\equiv& \sum_l \sum_{{i_1 \cdot\cdot\cdot i_{l-1}}} 
l ~ {\cal S}_{k, i_1,\cdot\cdot\cdot, i_{l-1}} ~ 
s^+_{i_1}\cdot\cdot\cdot s^+_{i_{l-1}}  ~~ ; \label{eq11} \\
G_{km}  &\equiv& \sum_l \sum_{{i_1 \cdot\cdot\cdot i_{l-2}}}
l(l-1) ~ {\cal S}_{k, m, i_1, 
\cdot\cdot\cdot, i_{l-2}} ~ s^+_{i_1} \cdot\cdot\cdot s^+_{i_{l-2}} ~~ .
\label{eq12}
\end{eqnarray}
The similarity transform of the Hamiltonian of Eq. (\ref{eq6}) 
may now be written in terms of these operators. We find explicitly
that $\hat H | \Phi \rangle \equiv e^{-S} H e^S | \Phi \rangle =
(\hat H_1 + \hat H_2 + \hat H_3) | \Phi \rangle$, where
\begin{eqnarray} 
\hat H_1 &\equiv&
-\frac 12 \sum_{i} \sum_{\rho}
\biggl\{
\Delta(G_{i,{i+\rho}}+ F_i F_{i+\rho}) + 
\frac 12 + G^2_{i,{i+\rho}}  
\nonumber \\
&& ~~~~~~~~~~~~~~~~
+ 2G_{i,{i+\rho}} F_i F_{i+\rho} + \frac 12  
F^2_i F^2_{i+\rho}
\biggr\}  s_i^+ s_{i+\rho}^+  ~~ ,
\label{eq13}   \\
\hat H_2 &\equiv&
\frac 12 \sum_{i} \sum_{\rho}
\biggl\{
\frac{\Delta}2 (F_{i+\rho}  s_{i+\rho}^+  +  F_i  s_i^+) 
+ (G_{i,{i+\rho}} + \frac 12  F_i F_{i+\rho})(F_{i+\rho} s_{i+\rho}^++ 
F_i  s_i^+)
\biggr\} ~~ ,
\label{eq14} \\ 
\hat H_3 &\equiv&
-\frac 14
\sum_{i} \sum_{\rho}
\biggl\{
\frac \Delta 2 + G_{i,{i+\rho}}+ F_i F_{i+\rho}
\biggr\} 
~~ .
\label{eq15}
\end{eqnarray}
We note that the similarity-transformed Hamiltonian acting on the
model state $| \Phi \rangle$ has now been written purely in terms 
of spin-raising operators. The problem of determining a given CCM 
ket-state equation, defined by Eq. (\ref{eq7}) for a given cluster 
of index $I$, thus becomes an exercise in pattern-matching 
the spin-lowering operators in $C_I^-$ to the terms in $\hat H$ 
contained in Eqs. (\ref{eq13}-\ref{eq15}). This task is perfectly 
suited to implementation using computer algebra techniques, and 
the ensuing set of coupled, non-linear ket-state equations 
is then easily solved (for example, using the Newton-Raphson 
method). Once the ket-state correlation coefficients have 
been determined it is then possible to find the bra-state 
coefficients similarly, as described in Ref.  \cite{ccm11}.
One may finally calculate any ground-state expectation 
values that one wishes to obtain in terms of the coefficients
$\{ {\cal S}_I, \tilde{{\cal S}_I} \}$. An example of this
is the sublattice magnetisation, given by
\begin{equation}
M = -\frac {2}N \langle \tilde \Psi \mid \sum_{i=1}^N s_i^z \mid \Psi 
\rangle ~~ ,
\label{eqmag}
\end{equation}
in the rotated spin coordinates defined above. 
The quantity provides a measure for the amount of N\'eel ordering in 
the $z$-direction remaining in the CCM ground state, $| \Psi \rangle$, 
with respect to the perfect ordering ($M=1$) of the model state.

The CCM formalism is exact in the limit of inclusion of
all possible multi-spin cluster correlations within 
$S$ and $\tilde S$, although in any real application 
this is usually impossible to achieve. It is therefore 
necessary to utilise various approximation schemes 
within $S$ and $\tilde{S}$. The three most commonly 
employed schemes have been: 
(1) the SUB$n$ scheme, in which all correlations 
involving only $n$ or fewer spins are retained, but no
further restriction is made concerning their spatial 
separation on the lattice; (2) the SUB$n$-$m$  
sub-approximation, in which all SUB$n$ correlations 
spanning a range of no more than $m$ adjacent lattice 
sites are retained; and (3) the localised LSUB$m$ scheme, 
in which all multi-spin correlations over distinct 
locales on the lattice defined by $m$ or fewer contiguous 
sites are retained. Note that for this system we also 
make the specific and explicit restriction that the 
creation operators $\{C_I^+\}$ in $S$ preserve the 
relationship that, in the original (unrotated) spin
coordinates,  $s^z_T=\sum_i s^z_i=0$ in order to 
keep the approximate CCM ground-state wave function in
the correct ($s^z_T=0$) subspace. The number of such 
distinct (or fundamental) configurations
for the ground state at a given level of approximation is 
labelled by $N_{F}$. We denote as distinct configurations
those which are inequivalent under the point and space
group symmetries of the both the lattice and the 
Hamiltonian.

In practice we find that the CCM equations often terminate at
specific critical values (denoted $\Delta_c$) of the anisotropy parameter
which are dependent on the particular approximation scheme chosen, 
such that no solution based on this model state exists for 
$\Delta < \Delta_c$. At these {\it critical points} the 
second derivative of the ground-state energy is found to diverge, 
and the critical points have been shown to reflect the 
corresponding phase transition point in the real system.

\subsection{The Excited-State Formalism}

We now turn our attention to the CCM parametrisation of the excited 
state developed by Emrich \cite{ccm15}, and as a specific example 
we present results for the high-order treatment of the excited states 
of the spin-half {\it XXZ} model. An excited-state wave function, 
$|\Psi_e\rangle$, is determined by linearly applying an excitation  
operator $X^e$ to the ket-state wave function of Eq. (\ref{eq2}), such that  
\begin{equation}
|\Psi_e\rangle = X^e ~ e^S |\Phi\rangle ~~ .
\label{eq16}
\end{equation}
This equation may now be used to determine the low-lying excitation 
energies, such that the Schr{\"o}dinger equation, $E_e |\Psi_e\rangle = 
H  |\Psi_e\rangle$, may be combined with its ground-state counterpart
of Eq. (\ref{eq1}) to give the result,
\begin{equation}
\epsilon_e X^e  | \Phi \rangle = e^{-S} [H,X^e] e^S | 
\Phi \rangle ~ (\equiv  \hat R | \Phi \rangle) ~~ ,
\label{eq17}
\end{equation}
where $\epsilon_e \equiv E_e-E_g$ is the excitation energy.
By analogy with the ground-state
formalism, the excited-state correlation operator is written as,
\begin{equation}
X^e = \sum_{I \ne 0} {\cal X}_I^e C_{I}^+ ~~ ,
\label{eq18}
\end{equation}
where the set $\{C_{I}^+\}$ of multi-spin creation 
operators may differ from those used in the ground-state 
parametrisation in Eq. (\ref{eq2}) if the excited 
state has different quantum numbers than the ground
state. 
We note that Eq. (\ref{eq18}) implies the overlap
relation $\langle \Phi | \Psi_e \rangle = 0$. By applying 
$\langle \Phi | C_I^-$ to Eq. (\ref{eq17}) we find that,
\begin{equation}
\epsilon_e {\cal X}_I^e = \langle \Phi | C_I^- e^{-S} [H,X^e] e^S | \Phi 
\rangle ~~ , \forall I \ne 0 ~~ , 
\label{temp1}
\end{equation}
which is a generalised set of eigenvalue equations with 
eigenvalues $\epsilon_e$
and corresponding eigenvectors ${\cal X}_I^e$, for each of the excited
states which satisfy $\langle \Phi | \Psi_e \rangle = 0$. Analogously to the 
ground-state case, we define the excited-state operators
\begin{eqnarray}
P_k & \equiv & \sum_l \sum_{i_1 \cdots i_{l-1}} 
l ~ {\cal X}_{k, i_1, \cdot\cdot\cdot
i_{l-1}}^e ~ s^+_{i_1}\cdot\cdot\cdot s^+_{i_{l-1}} ~~, \label{eq19} \\
Q_{km} & \equiv & \sum_l \sum_{i_1 \cdots i_{l-2}} 
l(l-1) ~ 
{\cal X}_{k,m, i_1, \cdot\cdot\cdot, i_{l-2}}^e 
~ s^+_{i_1} \cdot\cdot\cdot s^+_{i_{l-2}} ~~ . \label{eq20}
\end{eqnarray}
The state, $\hat R | \Phi \rangle \equiv e^{-S} [H,X^e] e^S | \Phi \rangle$, 
may now be divided into three elements, $\hat R | \Phi \rangle 
= (\hat R_1 + \hat R_2 + \hat R_3) | \Phi \rangle$, such that 
(after collecting together like terms) we find the following 
expressions,
\begin{eqnarray} 
\hat R_1 &\equiv&
-\sum_{i} \sum_{\rho}
\biggl\{
\Delta ( P_i F_{i+\rho} + \frac 12 Q_{i,{i+\rho}}) +
2  P_i F_{i+\rho} G_{i,{i+\rho}}  \nonumber \\
&& ~~~~~~~~~~~~ +
Q_{i,{i+\rho}} G_{i,{i+\rho}} +
Q_{i,{i+\rho}}  F_i F_{i+\rho}
+ P_i  F_i F_{i+\rho}^2 
\biggr\}  s_i^+ s_{i+\rho}^+ ~~ ,
\label{eq21}  \\
\hat R_2 &\equiv&
\sum_{i} \sum_{\rho}
\biggl\{
\frac {\Delta}2  P_i  s_i^+ +
 P_i G_{i,{i+\rho}}  s_i^+ +
 P_i  F_i F_{i+\rho}  s_i^+  
\nonumber \\
&& ~~~~~~~~~~ +
\frac 12  P_i F_{i+\rho} F_{i+\rho}  s_{i+\rho}^+ +
Q_{i,{i+\rho}} F_{i+\rho}  s_{i+\rho}^+
\biggr\} ~~ ,
\label{eq22} \\ 
\hat R_3 &\equiv&
-\frac 12
\sum_{i} \sum_{\rho}
\biggl\{
\frac 12 Q_{i,{i+\rho}}
+ P_i F_{i+\rho} 
\biggr\} 
~~ .
\label{eq23}
\end{eqnarray}
We note again that $\rho$ runs over all nearest-neighbour lattice vectors 
and that terms which are equivalent under a lattice translation which
connects one sublattice to the other have been collected together in Eqs. 
(\ref{eq21})-(\ref{eq23}). 
In order to determine explicitly the eigenvalue equation of Eq. 
(\ref{temp1}) we now {\it pattern match} the configurations in the 
set $\{C_{I}^{-}\}$ to the spin-raising operators contained
in $\hat R$, analogously to the ground-state procedure. 
For low orders of approximation, this may be performed readily 
by hand, although for higher orders of approximation may
one must again use computational methods.

We note that the lowest-lying excited states for the {\it XXZ} model 
lie in the $s_T^z=+1$ and $s_T^z=-1$ subspaces with respect to the 
``unrotated'' ground state. 
We thus restrict the ``fundamental'' clusters in the set $\{C_I^+\}$
used in Eq. (\ref{eq18}) to be those which reflect this property, 
and the number of such fundamental configurations for the excited 
state at a given level of LSUB$m$ approximation is denoted by $N_{F_e}$. 
In order to solve fully the eigenvalue problem 
at a given value of $\Delta$, we note furthermore that one 
must firstly fully determine and solve the ground ket-state 
equations in order to obtain numerical values for the set 
$\{ {\cal S}_I \}$ which are then used as input to the 
eigenvalue problem of Eq. (\ref{temp1}). 
The level of LSUB$m$ approximation is also explicitly restricted 
to be the same for the ground and excited states, such that
the calculation is kept as systematic and self-consistent as
possible. Finally, we note that for the specific case 
of the spin-half {\it XXZ} model considered here, we only 
consider the lowest eigenvalue of Eq. (\ref{eq17}) in order to 
calculate the excitation energy gap. We furthermore note that 
this eigenvalue is not specifically restricted by our truncation 
procedure to be a real number as the generalised eigenvalue 
problem of Eq. (\ref{eq17}) is not constrained to be symmetric. 
However, in practice we find that this eigenvalue is real at 
every value of $\Delta$ considered, for each lattice, and at 
each level of LSUB$m$ approximation. This provides a rather 
stringent check on the approximation scheme. 

\subsection{Extrapolation of CCM Results}

We present results below of high-order CCM calculations for the 
spin-half {\it XXZ} model for the linear chain, square lattice, 
and simple cubic lattice. In particular, we determine the 
ground-state energy, the sublattice magnetisation, and the 
lowest-lying excited-state energy for these models using the 
LSUB$m$ approximation scheme. It is clearly useful to be able 
to extrapolate the ``raw'' CCM results, at each value of 
$\Delta$ separately, to the limit $m \rightarrow \infty$.
In the absence of any rigorous scaling theory for the results of
LSUB$m$ approximations, two empirical approaches are outlined
below and utilised. The first approach, denoted 
to as ``Extrapolated(1) CCM,'' assumes a leading ``power-law'' 
dependence of the LSUB$m$ expectation values with $m$. The value
for the ground-state energy, sublattice magnetisation, or excitation
energy (at a given value of $\Delta$) is denoted by $y_i$ for a given
value of $m$ where $x_i \equiv 1/m$. The index $i$ denotes the i$^{th}$ 
data element of $p$ such elements, although we note that $i$ 
and $m$ do not have to be explicitly equivalent. The leading 
``power-law'' extrapolation is now described by,
\begin{eqnarray}
y_i &=& a + b x_i^{\nu} ~~ .
\nonumber 
\end{eqnarray}
We plot ${\rm log}(x_i)$ against ${\rm log}(y_i-a)$ 
and the best fit of the 
data set, $\{x_i,y_i\}$, to the power-law dependence given above 
is obtained when the absolute value of the {\it linear correlation}
of these points is maximised with respect to the variable $a$. This
value of $a$ is then assumed to be the extrapolated value for 
$y_i$ in the limit $m \rightarrow \infty$.

The second such extrapolation scheme of the LSUB$m$ data, 
denoted the ``Extrapolated(2) CCM'' scheme, uses 
Pad\'e approximants in which the data set is modeled by the ratio of 
two polynomials, given by
\begin{eqnarray}
y_i &=& \frac {\sum_{j=0}^k a_j x_i^j} {1 + \sum_{j=1}^l b_j x_i^j} ~~ .
\nonumber 
\end{eqnarray}
This furthermore implies that,
\begin{eqnarray}
a_0 + a_1x_i + a_2 x_i^2 + \cdots + a_k x_i^k
&=& y_i +(b_1x_i + b_2x_i^2 + \cdots + b_l x_i^l) y_i ~~.
\nonumber 
\end{eqnarray}
We now wish to determine 
the coefficients $a_j$ and $b_j$ in order to find the polynomials in
the equation above, and the above equation is rewritten in terms of a 
matrix given by,
\begin{equation}
\left (
\mbox{
\begin{tabular}{ccccc|cccc}
1, &$x_1$, &$x_1^2$, &$\cdots$, &$x_1^k$  &$x_1y_1$, &$x_1^2y_1$, &$\cdots$, 
&$x_1^ly_1$ \\
1, &$x_2$, &$x_2^2$, &$\cdots$, &$x_2^k$  &$x_2y_2$, &$x_2^2y_2$, &$\cdots$, 
&$x_2^ly_2$ \\
   &$\cdot$&         &$\cdot$   &         &          &$\cdot$     &$\cdot$
&         \\ 
   &$\cdot$&         &$\cdot$   &         &          &$\cdot$     &$\cdot$
&         \\ 
   &$\cdot$&         &$\cdot$   &         &          &$\cdot$     &$\cdot$
&         \\ 
   &$\cdot$&         &$\cdot$   &         &          &$\cdot$     &$\cdot$
&         \\ 
1, &$x_p$, &$x_p^2$, &$\cdots$, &$x_p^k$  &$x_p y_p$, &$x_p^2y_p$, &$\cdots$, 
&$x_p^ly_p$ \\
\end{tabular}
}
\right ) 
\left (
\mbox{
\begin{tabular}{c}
$a_0$   \\
$a_1$   \\
$\cdot$ \\
$a_k$   \\
$-b_1$   \\
$\cdot$ \\
$-b_l$   \\
\end{tabular}
}
\right ) =
\left (
\mbox{
\begin{tabular}{c}
$y_1$ \\
$y_2$ \\
$\cdot$ \\
$\cdot$ \\
$\cdot$ \\
$\cdot$ \\
$y_p$ \\
\end{tabular}
}
\right )
\label{eq27}
\end{equation}
(Note that $k+l+1=p$.) This problem may be easily solved by 
inverting the matrix in order to determine the coefficients
$\{a_i\}$ and $\{b_i\}$ at each value of $\Delta$ separately.
In the limit, $m \rightarrow \infty$, it is seen that the 
extrapolated value of $y_i$ is given by $a_0$. We furthermore
note that the case $l=0$ corresponds also to a ``least-squares''
fit of the $p$ data points to a $(p-1)^{\rm th}$-order polynomial.



Previous extrapolation results  \cite{ccm11} for the HAF on the square 
lattice fitted $y_i$ against $x_i=1/m^2$ for the ground-state energy, 
$E_g/N$, and the critical points, $\Delta_c$, while the sublattice 
magnetisation was fitted against $x_i=1/m$. 
These ``rules'' were chosen because the CCM SUB2-$n$ calculations for 
the square lattice HAF were found to converge as a function of $n$ to 
their full SUB2 solution in these ways, and so analogous rules were 
used for the LSUB$m$ data.
Note that results for the HAF on the square and simple cubic lattices using 
these procedures are also quoted below. For the spin-half square 
lattice HAF, we find that the present rules are reasonable 
approximations for the scaling of the LSUB$m$ results.

Finally, we note that the LSUB2 results for $y_i$ and $x_i$ 
generally fit in rather poorly to the asymptotic behaviour of 
$y_i$ and $x_i$ as a function of $i$. As we are interested in the 
asymptotic value of $y_i$, it is found that we obtain better 
extrapolated results by discarding the lowest order LSUB2 results 
where possible, i.e., for the linear chain and square lattice 
results. Hence for both extrapolation schemes used here, LSUB$m$ 
results are used with $m=\{4,6,8,10,12\}$ for the linear chain,
with $m=\{4,6,8\}$ for the square lattice, but with $m=\{2,4,6\}$ 
for the simple cubic lattice. 

\section{Results}

\begin{figure}
\epsfxsize=14cm
\centerline{\epsffile{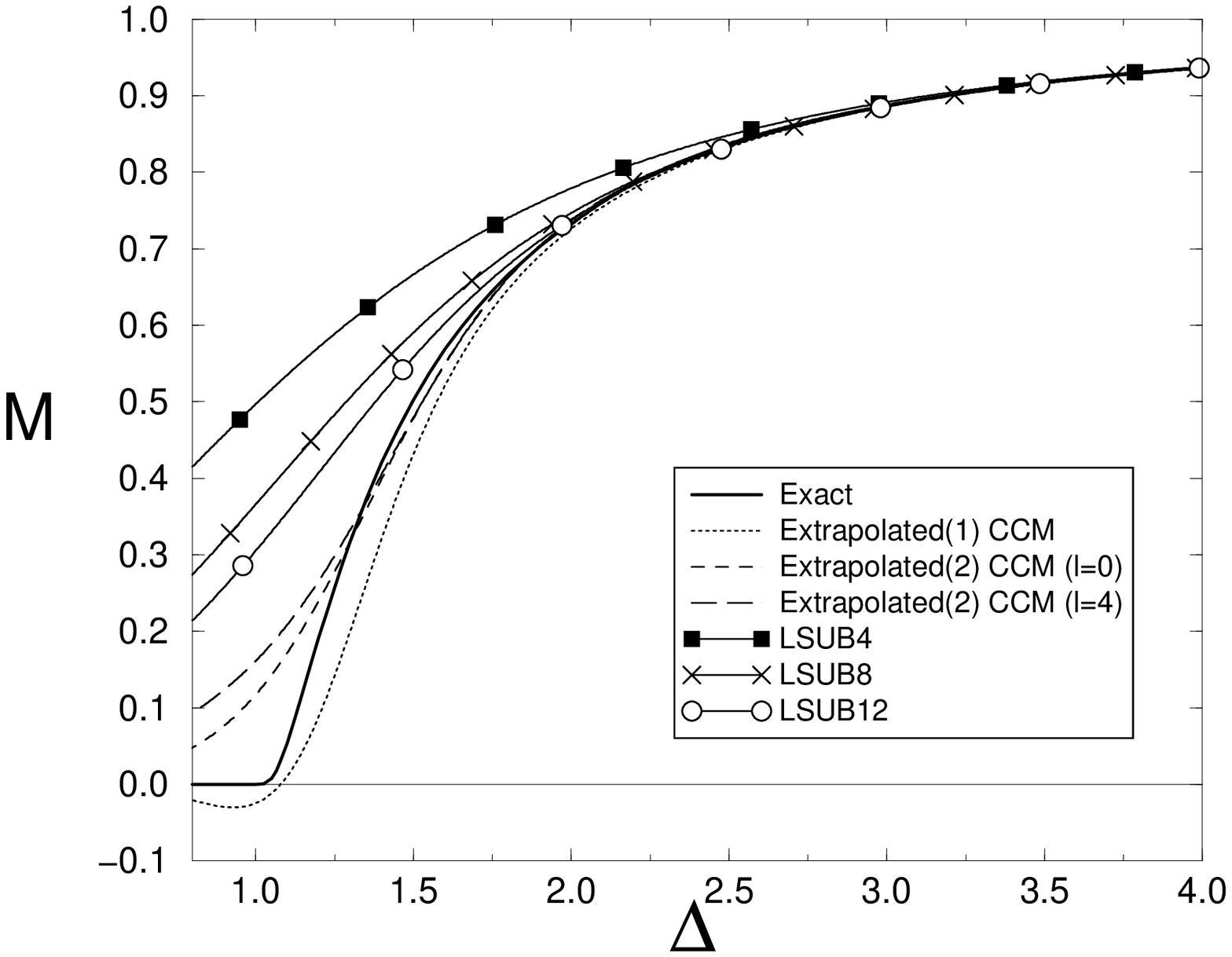}}
\vspace{0.4cm}
\caption{}
\label{fig1}
\end{figure}

\begin{figure}
\epsfxsize=14cm
\centerline{\epsffile{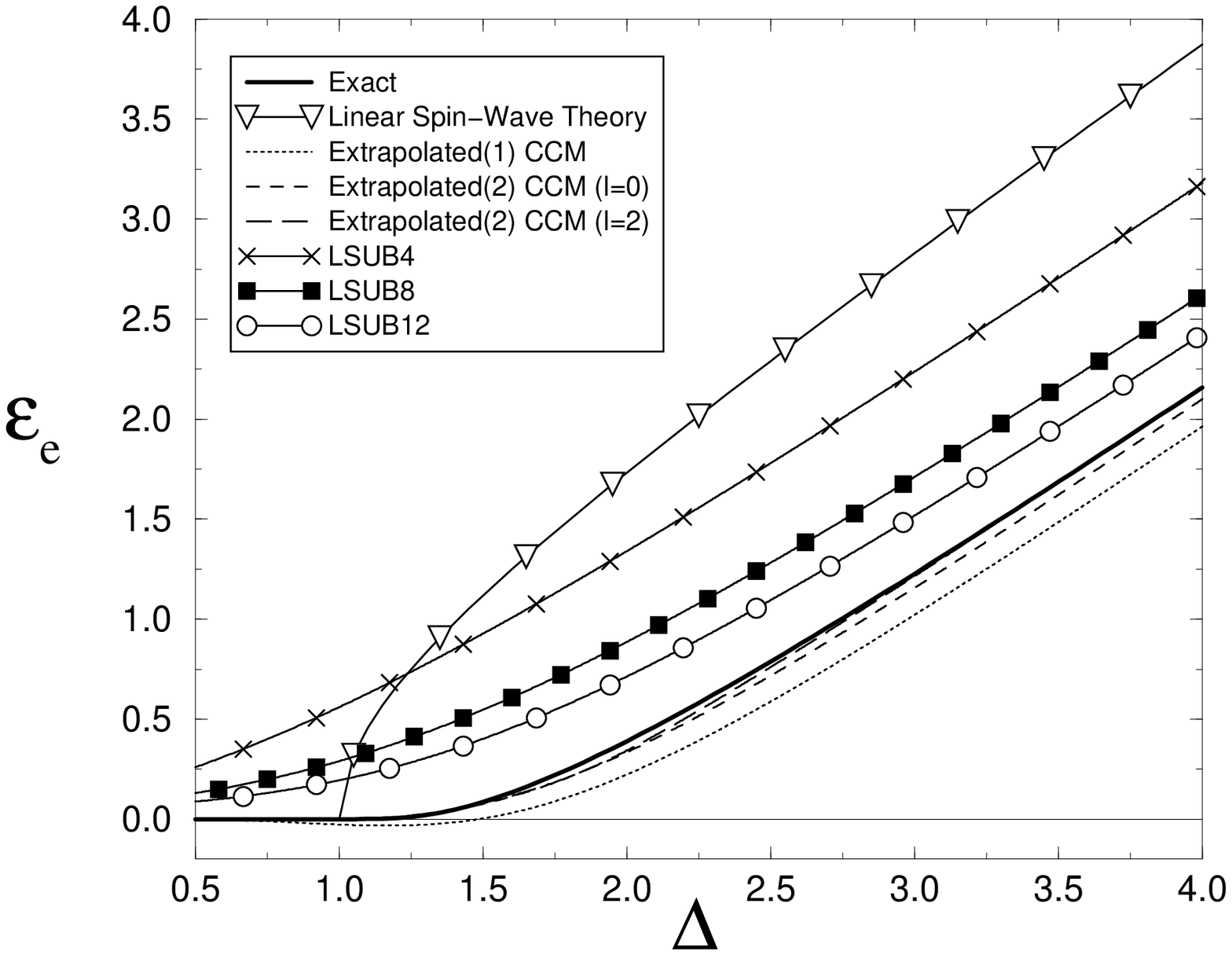}}
\vspace{0.4cm}
\caption{}
\label{fig2}
\end{figure}

The results for the ground-state energy, $E_g/N$, of the CCM 
treatment of the spin-half linear chain {\it XXZ} model regime 
$\Delta \ge 1$ are found to be in excellent agreement with 
exact results \cite{ba1,ba2,ba3,ba4} over the whole of this regime, 
(and therefore no graphical plot of this data is given). Table 
\ref{tab1} indicates the accuracy of the raw and extrapolated 
LSUB$m$ for the spin-half linear chain HAF ($\Delta=1$) in 
comparison with exact Bethe Ansatz  \cite{ba1,ba2,ba3,ba4} 
results. It is seen that all of the extrapolated results for
the ground-state energy, apart from those of the extrapolated(2) 
CCM scheme with $l=3$, are accurate to 
at least five decimal places. Indeed, the spin-half linear chain 
HAF is the ``worst-case-scenario'' within this regime for 
such CCM calculations. Thus, for $\Delta > 1$ even better accuracy 
is obtained, and indeed the CCM results are exact at all levels of 
approximation in the limit $\Delta \rightarrow \infty$, in which 
case the model state is the exact quantum ground state.
The results for the sublattice magnetisation, $M$, of this 
model using the CCM are shown in Fig. \ref{fig1}. It is seen
that the CCM results tend to the exact results as $\Delta$ 
increases, although they are non-zero and positive at 
$\Delta=1$. The raw CCM results may be greatly improved 
by extrapolation, and the extrapolated(1) CCM results of
$M=-0.0247$ is in good agreement with the exact result
that $M$ becomes zero at $\Delta=1$. The
extrapolated(2) CCM results do not perform quite as well
as the extrapolated(1) CCM results, although they do present
a very significant improvement on the raw LSUB$m$ results.
However, Fig. \ref{fig1} shows that the results of both
extrapolations are in good qualitative agreement with exact
results, and we note that the spin-half HAF on the
linear chain presents a particularly difficult challenge,
since not only does all of the N\'eel-like long-range order inherent 
in the model state disappear at this point but also there
is an infinite-order phase transition at this point (i.e., 
one in which the energy and all of its derivatives with
respect to $\Delta$ are continuous). An 
indication of the efficacy of the CCM treatment of this 
model is also given for the lowest-lying excitation energy,
as shown in Fig. \ref{fig2}. It is seen that the extrapolated 
results are in excellent agreement with the exact result, 
and are very much better than those results of linear 
spin-wave theory (LSWT) 
over the whole of this regime. Indeed, the manner in which 
the CCM results for the excitation energy behave as $\Delta 
\rightarrow 1$ is in good qualitative and quantitative 
agreement with the exact results. This is in stark contrast 
to the results of LSWT in 1D which clearly show completely
incorrect qualitative behaviour for $\epsilon_e$ near 
to $\Delta=1$. Furthermore, Table \ref{tab1} indicates 
that the extrapolated(2) CCM results for the excitation
energy of the spin-half linear chain HAF are zero to three 
decimal places of accuracy. 
The extrapolated(1) results are also consistent with a 
gapless spectrum at $\Delta=1$. 

\begin{figure}
\epsfxsize=14cm
\centerline{\epsffile{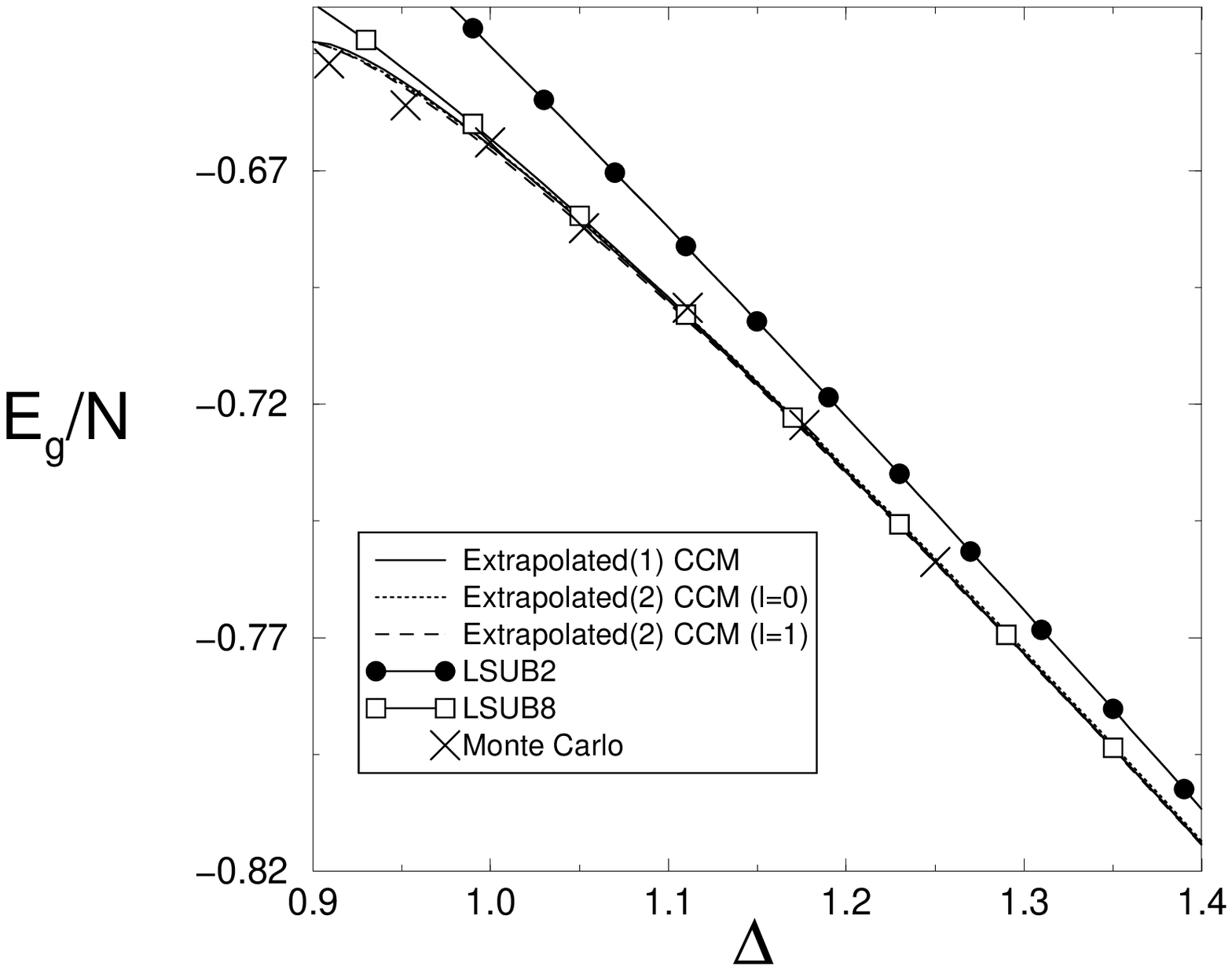}}
\vspace{0.4cm}
\caption{}
\label{fig3}
\end{figure}

The various CCM results for the ground-state energy 
of the spin-half square lattice {\it XXZ} model are found to 
be in excellent mutual agreement over a wide range
of $\Delta$. This may be seen in Fig. \ref{fig3}, in which both
the ``raw'' ground-state energy CCM LSUB$m$ results for 
$m=\{2,4,6,8\}$ and extrapolated CCM results are plotted for 
this model. It is also seen from Fig. \ref{fig3} that 
our results are in excellent agreement with those results 
of the QMC method given in Ref.  \cite{qmc1} for this model.
Table \ref{tab2} illustrates the CCM results for the 
special case of the spin-half HAF, and it is seen 
that CCM results are in excellent agreement (see also 
for example Refs.  \cite{ccm2,ccm7,ccm11}) with those 
results of SWT \cite{swt1,swt2}, finite-sized 
calculations  \cite{finite1}, series expansions
 \cite{series1}, and QMC techniques  \cite{qmc1,qmc2,qmc3,qmc4}. 
However, our results for the spin-half square lattice 
HAF perhaps indicate that the ground-state energy might 
be slightly lower than the best QMC estimate obtained
to date of Sandvik \cite{qmc4} which is, namely, $E_g/N = 
-0.669437(5)$. The extrapolated(1) CCM and extrapolated(2)
CCM results are in also good agreement with the previous
extrapolation  \cite{ccm11} of CCM LSUB$m$ results at
this point, which predict that $E_g/N = -0.66968$. 
(We note again that the previous 
extrapolated value  \cite{ccm11} of $E_g/N = -0.66968$ 
was obtained by fitting to a quadratic in $1/m^2$.) 

As $\Delta$ approaches the HAF ($\Delta=1$), 
we find that the ground-state energy exponent, $\nu$,
approaches the value $\nu \approx 2$. 
We also note that the expansion coefficient of the $l=0$ Pad\'e 
approximant (viz., the series in integral powers of 
$1/m$) which becomes overwhelmingly dominant as $\Delta 
\rightarrow 1$ is the coefficient which corresponds 
to the $1/m^2$ term. Thus, the extrapolated(1) and 
extrapolated(2) results have independently reinforced 
what were the admittedly ``naive'' assumptions made 
in the ``previous'' extrapolation  \cite{ccm11} 
in which the ground-state energy for the square 
lattice HAF is fitted to a quadratic function in 
$1/m^2$. (Note that similar results for the scaling 
of the ground-state energy are also observed for 
both the linear chain and simple cubic lattice.)

\begin{figure}
\epsfxsize=14cm
\centerline{\epsffile{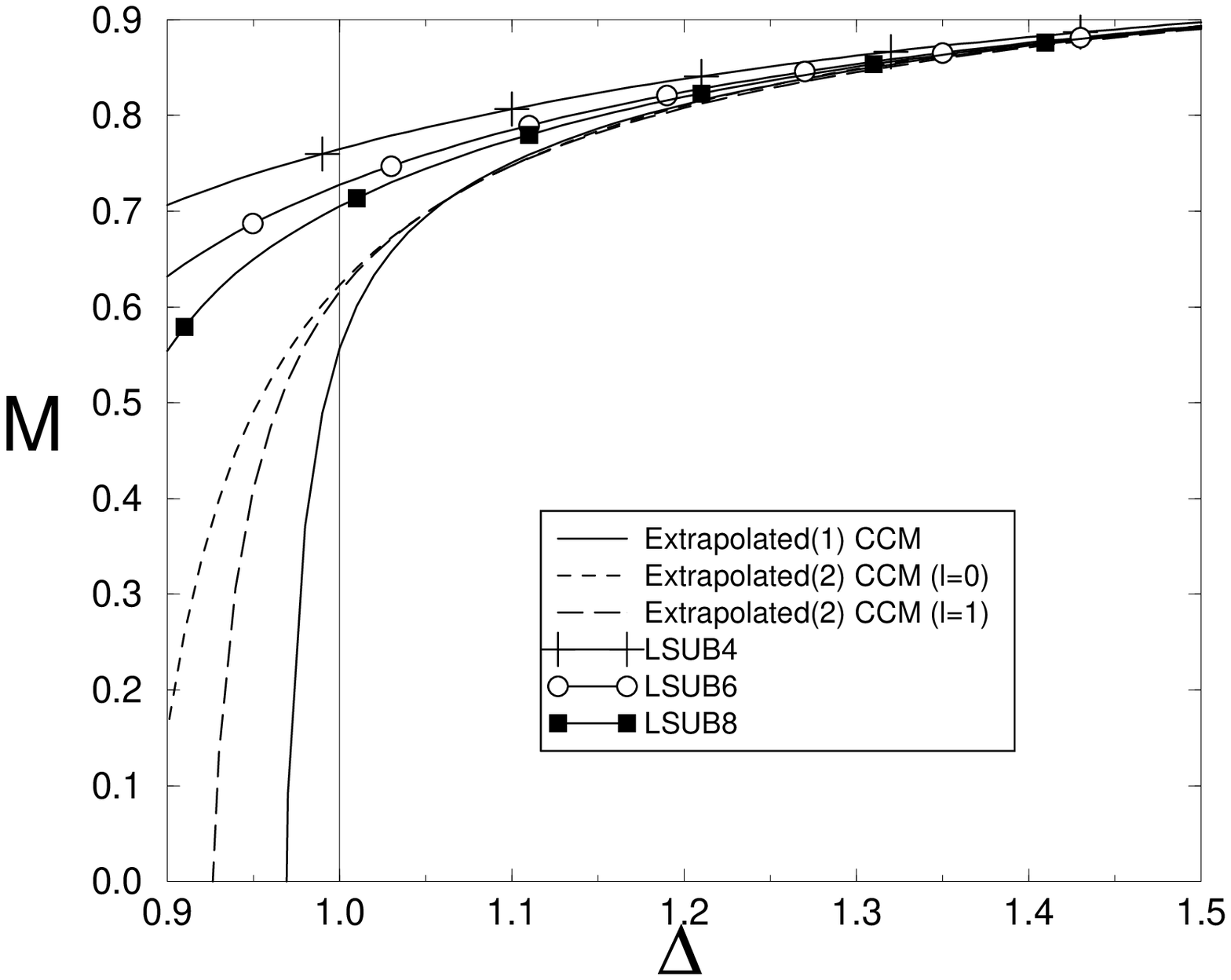}}
\vspace{0.4cm}
\caption{}
\label{fig4}
\end{figure}

The various CCM results for the sublattice magnetisation 
of the spin-half square lattice {\it XXZ} model are similarly 
in very good mutual agreement over a wide range
of $\Delta$, as seen in Fig. \ref{fig4}. 
They are furthermore found to agree very well 
with the results of the best of other approximate
methods for the specific case of the HAF 
at $\Delta = 1$, as shown in Table \ref{tab2}, 
which typically yield results of $M \approx 0.61$. 
However, we note that the extrapolated(1) 
CCM result for the HAF of $M=0.557$ given in Table 
\ref{tab2} probably lies slightly too low. 
This is because the critical exponent $\nu$ for the sublattice
magnetisation appears to change very rapidly near to 
the HAF point for the square lattice,
where, for example, we find that $\nu=1$ at $\Delta=1.06$ 
and that $\nu = 0.490$ at $\Delta=1$. 
It is expected however that the inclusion of the second-order 
term to the extrapolation using a leading-order power-law 
scaling would remedy this situation. 
We may see from Fig. \ref{fig4}, however, that the 
extrapolated(2) CCM results appear to work extremely well
up to and including the HAF itself. 
(Note that qualitatively similar results are observed 
for the simple cubic lattice as for the square lattice.)

\begin{figure}
\epsfxsize=14cm
\centerline{\epsffile{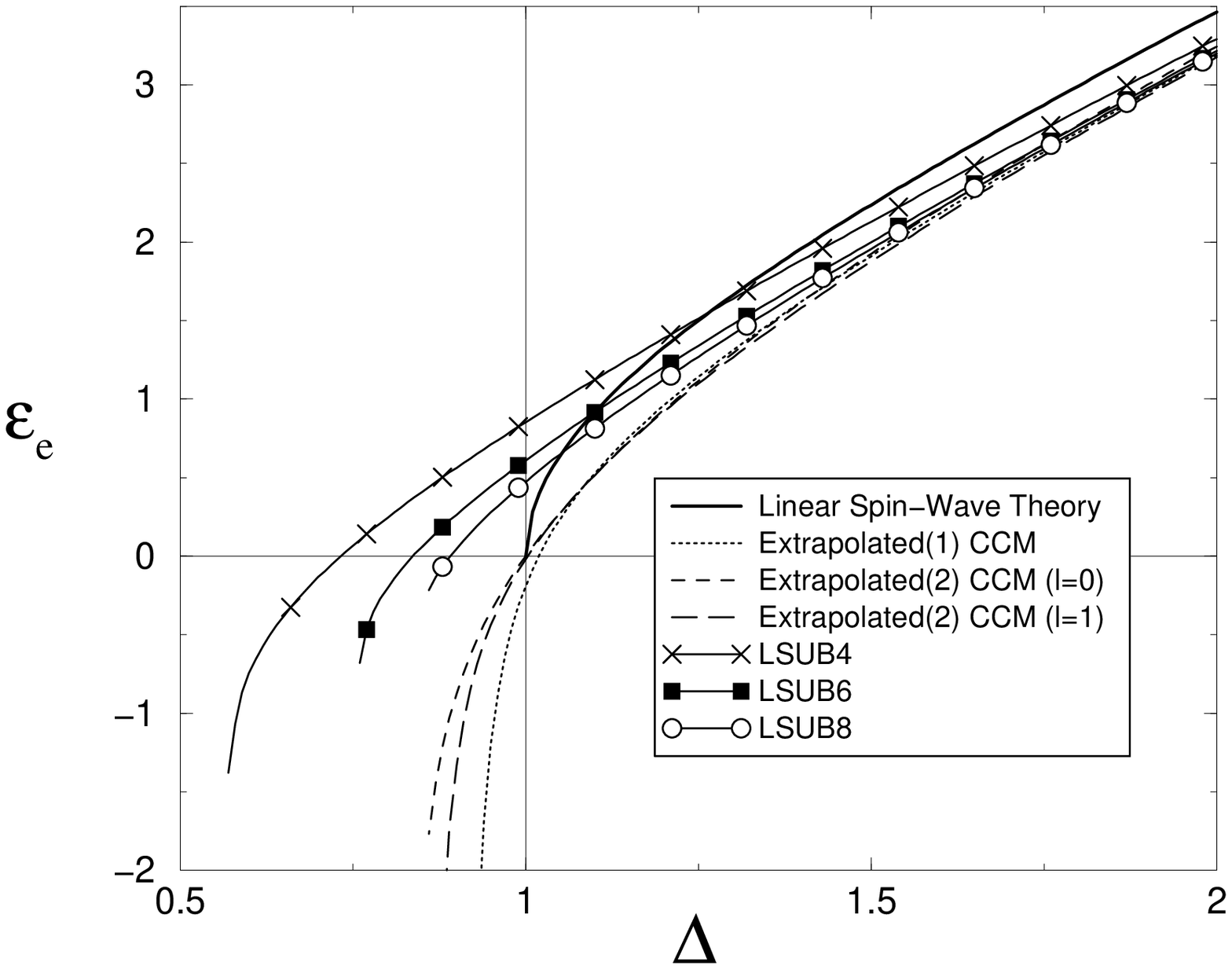}}
\vspace{0.4cm}
\caption{}
\label{fig5}
\end{figure}

The CCM results for the lowest-lying excitation energy of the 
{\it XXZ} model in the $\Delta > 1$ regime also appear to be 
in excellent mutual agreement, as shown in Fig. \ref{fig5}. 
The extrapolated CCM results seem to be qualitatively similar 
to results of LSWT, although they lie significantly lower and 
are believed to be much more accurate than the LSWT results 
in this regime. The mutual agreement of the extrapolated CCM 
results only ever falters very near to the HAF point
($\Delta=1$). Indeed, the spin-half square lattice HAF 
is believed to be gapless. The extrapolated(1) CCM 
results predict that the excitation energy gap disappears at 
$\Delta=1.02$. Furthermore, the extrapolated(2) CCM $l=0$ and 
$l=1$ results strongly indicate that the spin-half square lattice 
isotropic HAF ($\Delta=1$) is gapless, with results of 
$\epsilon_e=-0.001$ and $\epsilon_e=-0.020$, respectively. 
Indeed, we believe that the extrapolated(2) CCM 
results provide excellent results for the excitation 
gap of the spin-half square lattice {\it XXZ} model across 
the whole of the regime, $\Delta \ge 1$. 

The results for the critical points predicted by the CCM for 
the spin-half square  lattice {\it XXZ} model are also presented
in Table \ref{tab2}. These results strongly indicate that the 
phase transition point is at (or very near to) the HAF 
point ($\Delta=1$), which is fully consistent with the
behaviour that is believed to occur in this model.

We have also determined completely new CCM results for the spin-half 
{\it XXZ} model on the simple cubic lattice. These results 
look qualitatively very similar to those for the square lattice
model and so are not plotted here. It is seen from Table \ref{tab3} 
that the results for the ground- and excited-state properties of 
the cubic-lattice HAF are in good agreement with those results 
of LSWT. The results for ground-state energy of the
spin-half cubic-lattice HAF are shown in Table \ref{tab3}, and 
it is seen that the LSUB6 approximation is already quite 
near the converged limit. The extrapolated(1) and extrapolated(2) 
CCM results and the ``previous extrapolated CCM'' result for the 
ground-state energy are thus seen to be in mutual agreement
with each other. For the sublattice magnetisation
of the spin-half cubic-lattice HAF, the extrapolated 
CCM results give values of $M = 0.836\pm0.001$, and this
result is in excellent agreement with the result of LSWT
of $M = 0.843$. By contrast, the extrapolated(1) CCM result appears 
to lie too low for the lowest-lying excitation energy of the 
simple cubic-lattice HAF. However, the extrapolated(2) CCM $l=0$ and 
$l=1$ results indicate that the spin-half cubic-lattice HAF 
is gapless with results of $\epsilon_e=0.008$ and $\epsilon_e 
= -0.057$, respectively, which is once again consistent 
with the expected behaviour of this model.

The results for the critical point of the spin-half cubic-lattice
{\it XXZ} model appear to converge to a value near to $\Delta=1$, 
which is again believed to be (or be near to) the correct result. 
(Note that with only two points an assumption must be made as to 
the scaling of $\Delta_c$ with $m$. Hence, in analogy with SUB2-$n$ 
results the LSUB$m$ values for $\Delta_c$ were plotted against 
$1/m^2$ and linearly extrapolated, and so these results are referred
to as the ``previous extrapolated CCM'' result in Table \ref{tab3}.)

\section{Conclusions}





In this article new results have been presented for the ground-state
properties of the spin-half {\it XXZ} model on the linear chain, the
square lattice, and the simple cubic lattice. A localised approximation scheme
was utilised within the CCM ground-state formalism, and this 
allowed high-order calculations to be very efficiently performed. 
The results were seen to be in excellent agreement with exact Bethe 
Ansatz calculations for the linear chain and the best results of 
other approximate methods for the square and simple cubic lattices. These 
results were extrapolated using a number of techniques which extend 
previous heuristic attempts at extrapolation. The results so
obtained are of very high quality across a very wide
range of the anisotropy parameter $\Delta$, thus reinforcing our
results at the specific point for the Heisenberg antiferromagnet at 
$\Delta=1$. Indeed, it was found from these calculations that our 
previous heuristic extrapolations for the ground-state energy
of the spin-half square lattice Heisenberg antiferromagnet 
did, in fact, provide very reasonable results, and that the 
naive assumptions made in this extrapolation were 
largely justified. Hence, for example, the previous 
``best estimate'' of the ground-state energy of the spin-half
square-lattice Heisenberg antiferromagnet of $E_g/N = 
-0.66968$ was justified. Furthermore, our CCM 
calculations for spin-half Heisenberg antiferromagnet 
indicate that about 62$\%$ of the classical N\'eel ordering
remains for the square lattice case, and that 83.6$\%$ of the 
classical N\'eel ordering remains for the simple cubic lattice case.

A new formalism has also been introduced in order to treat
for the first time the excited states of lattice quantum 
spin systems via the same high-order localised approximation 
scheme in a computationally efficient manner. 
The approximation level was kept at the same subapproximation
level for both the ground and excited states, although
the CCM ground-state correlations preserved the $s_T^z =0$
property of the ground state and the CCM excited-state 
correlations produced a change of $s_T^z = \pm 1$ with 
respect to the CCM approximation to the ket ground state. 
Extrapolations were again attempted in the $\Delta \ge 1$ regime
and very good correspondence of the extrapolated CCM
results for the lowest-lying excitation energy were 
observed for the linear chain in comparison with the
exact Bethe Ansatz results. Furthermore, the results for 
the spin-half Heisenberg antiferromagnet, for all of the 
lattices considered here, strongly indicate that the 
excitation energy gaps of these models are zero. 

Future applications of the high-order formalism presented
in this article will be to systems of complex crystallographic
point- and space-group symmetries, e.g., those of the 
CaVO. Another extension will 
be to apply these methods to systems of higher quantum
spin number, $s$. In particular, it would be of interest
to apply the new high-order formalism for the excited
state to the spin-one linear chain Heisenberg antiferromagnet 
which contains an excitation energy gap. The high-order
formalism might also be applied to systems with electronic 
(rather than just spin) degrees of freedom. Finally, the 
application of the CCM to spin systems at non-zero temperature 
remains a future goal.

\section*{Acknowledgements}

We wish to thank Dr. N.E. Ligterink for his useful and interesting 
discussions. 
One of us (RFB) gratefully acknowledges a research grant from the
Engineering and Physical Sciences Research Council (EPSRC) of Great
Britain. This work has also been supported by the Deutsche
Forschungsgemeinschaft (GRK 14, Graduiertenkolleg on `Classification of 
Phase Transitions in Crystalline Materials' and also 
Project No. RI 615/9-1).  One of us (RFB) also acknowledges the support 
of the Isaac Newton Institute for Mathematical Sciences, 
University of Cambridge, during a stay at which the
final version of this paper was written.

\pagebreak


\begin{table}
\caption{Results obtained for the spin-half HAF on the 
linear chain using the CCM LSUB$m$ approximation with $m=\{2,4,6,8,10,12\}$. 
$N_{F}$ denotes the number of fundamental configurations for the 
ground state, and $N_{F_e}$ denotes the number of fundamental 
configurations for the excited state. The ground-state energy per 
spin, $E_g/N$, the sublattice magnetisation, $M^+$, and the 
lowest-lying excitation energy, $\epsilon_e$, are shown. The
LSUB$m$ results for $m=\{4,6,8,10,12\}$ are extrapolated using the
two extrapolation schemes described in the text, and these results 
are compared to exact results from the Bethe Ansatz.}
\begin{center}
\begin{tabular}{|l|c|c|c|c|c|}  \hline\hline
        &$N_{F}$    	&$N_{F_e}$    	 &$E_g/N$ 
        &$M^+$          &$\epsilon_e$    \\ \hline\hline
LSUB2   &1              &1               &$-$0.416667 
        &0.666667       &1.000000      \\ \hline
LSUB4   &3              &3               &$-$0.436270
        &0.496776       &0.560399      \\ \hline
LSUB6   &9              &9               &$-$0.440024
        &0.415771       &0.383459      \\ \hline
LSUB8   &26             &31              &$-$0.441366
        &0.365943       &0.290251      \\ \hline
LSUB10  &81             &110             &$-$0.441995  
        &0.331249       &0.233098      \\ \hline
LSUB12  &267            &406             &$-$0.442340
        &0.305254       &0.194577      \\ \hline
Extrapolated(1) CCM 
        &--             &--  		 &$-$0.443152
	&$-$0.0247      &$-$0.0270   \\ \hline
Extrapolated(2) CCM ($l=0$) &-- &-- 	 &$-$0.443154	
	&0.1351 	&$-$0.0009   \\ \hline
Extrapolated(2) CCM ($l=1$) &-- &-- 	 &$-$0.443152
	&0.1043         &$-$0.0006      \\ \hline
Extrapolated(2) CCM ($l=2$) &-- &-- 	 &$-$0.443153
	&0.0938         &$-$0.0008      \\ \hline
Extrapolated(2) CCM ($l=3$) &-- &-- 	 &$-$0.443171
	&0.1128         &$-$0.0009   \\ \hline
Bethe Ansatz [23-26]
        &--             &--         &$-$0.443147    
        &0.0            &0.0             \\ \hline
\end{tabular}
\end{center}
\vspace{20pt}
\label{tab1}
\end{table}

\begin{table}
\caption{Results obtained for the spin-half HAF on the 
square lattice using the CCM LSUB$m$ approximation with $m=\{2,4,6,8\}$. 
$N_{F}$ denotes the number of fundamental configurations for the 
ground state, and $N_{F_e}$ denotes the number of fundamental 
configurations for the excited state. The ground-state energy per 
spin, $E_g/N$, the sublattice magnetisation, $M^+$, the 
lowest-lying excitation energy, $\epsilon_e$, and the critical
points, $\Delta_c$, are shown. The LSUB$m$ results $m=\{ 4,6,8\}$ 
are extrapolated using the two extrapolation schemes described in 
the text, and these results are compared to the best results of 
other approximate methods.} 
\begin{center}
\begin{tabular}{|l|c|c|c|c|c|c|}  \hline\hline
        &$N_{F}$    	&$N_{F_e}$    	 &$E_g/N$ 
        &$M^+$          &$\epsilon_e$    &$\Delta_c$\\ \hline\hline
LSUB2   &1              &1               &$-$0.648331
        &0.841427       &1.406674        &--    \\ \hline
LSUB4   &7             &6               &$-$0.663664
        &0.764800       &0.851867        &0.577 \\ \hline
LSUB6   &75             &91              &$-$0.667001
        &0.727282       &0.609657        &0.763 \\ \hline
LSUB8   &1273           &2011            &$-$0.668174
        &0.704842       &0.472748        &0.843 \\ \hline
Extrapolated(1) CCM 
        &--             &--  		 &$-$0.669695
	&0.557          &$-$0.191        &1.001 \\ \hline
Extrapolated(2) CCM ($l=0$)  
        &--             &-- 	         &$-$0.669713
	&0.623   	&$-$0.001        &1.031 \\ \hline
Extrapolated(2) CCM ($l=1$)  
        &--             &-- 	         &$-$0.670619
	&0.616          &$-$0.020        &1.044 \\ \hline
Previous Extrapolated CCM [19]
        &--             &--              &$-$0.66968	
        &0.62	        &--              &0.96  \\ \hline
Linear SWT [27]   
        &--             &--              &$-$0.65795
        &0.6068         &0.0             &1.0   \\ \hline
Second-order SWT [28]
        &--             &--              &$-$0.67042
        &0.6068         &0.0             &1.0   \\ \hline
Third-order SWT [28]
        &--             &--              &$-$0.66999
        &0.6138         &0.0             &1.0   \\ \hline
Series Expansions [31]
        &--             &--              &$-$0.6693(1)
        &0.614(2)       &--              &--    \\ \hline
QMC, Barnes {\it et al.} [32]
        &--             &--              &$-$0.669
        &--             &--              &--    \\ \hline
QMC, Runge [34]
        &--             &--              &$-$0.66934(4)
        &0.615(5)       &--              &--    \\ \hline
QMC, Sandvik [35]
        &--             &--              &$-$0.669437(5)
        &0.6140(6)      &--              &--    \\ \hline
\end{tabular}
\end{center}
\vspace{20pt}
\label{tab2}
\end{table}

\begin{table}
\caption{Results obtained for the spin-half HAF on the simple 
cubic lattice using the CCM LSUB$m$ approximation with $m=\{2,4,6\}$.
$N_{F}$ denotes the number of fundamental configurations for the 
ground state, and $N_{F_e}$ denotes the number of fundamental 
configurations for the excited state. The ground-state energy per 
spin, $E_g/N$, the sublattice magnetisation, $M^+$, the 
lowest-lying excitation energy, $\epsilon_e$, and the critical
points, $\Delta_c$, are shown. The LSUB$m$ results for 
$m=\{2,4,6\}$ are extrapolated using the two extrapolation 
schemes described in the text, and these
results are compared to linear spin-wave theory (SWT)
results of Ref. [27]. The results denoted as 
``previous extrapolated CCM'' (see Sec. III) 
use the $1/m^2$ extrapolation ``rule'' for the ground-state energy 
and critical point in order to extrapolate the raw 
LSUB$m$ result. However, we note that they have 
not previously been published for the spin-half cubic 
lattice cases.}
\begin{center}
\begin{tabular}{|l|c|c|c|c|c|c|}  \hline\hline
        &$N_{F}$    	&$N_{F_e}$    	&$E_g/N$ 
        &$M^+$          &$\epsilon_e$   &$\Delta_c$ \\ \hline\hline
LSUB2   &1              &1              &$-$0.890755
        &0.900472       &1.873964       &--         \\ \hline
LSUB4   &8              &7              &$-$0.900434
        &0.867849       &1.117651       &0.690      \\ \hline
LSUB6   &181            &223            &$-$0.901802
        &0.857192       &0.787066       &0.843      \\ \hline
Extrapolated(1) CCM 
        &--             &--             &$-$0.9026
	&0.837          &$-$0.610       &--         \\ \hline
Extrapolated(2) CCM ($l=0$)  
        &--             &--             &$-$0.9018
	&0.836          &0.008          &--         \\ \hline
Extrapolated(2) CCM ($l=1$)  
        &--             &--    	        &$-$0.9036
	&0.836          &$-$0.057       &--         \\ \hline
Previous Extrapolated CCM 
        &--             &--             &$-$0.9028 
        &--             &--             &0.965      \\ \hline
Linear SWT [27]
        &--             &--             &$-$0.896
        &0.843          &0.0            &1.0        \\ \hline 
\end{tabular}
\end{center}
\vspace{20pt}
\label{tab3}
\end{table}

\pagebreak

\section*{Figure Captions}

\vspace{0.5cm}

\noindent
{\bf Figure 1:} CCM results for the sublattice magnetisation of the spin-half 
{\it XXZ} model on the linear chain using the LSUB$m$ approximation
based on the $z$-aligned N\'{e}el model state, compared to exact Bethe 
Ansatz results.

\vspace{0.5cm}

\noindent
{\bf Figure 2:} CCM results for the lowest-lying excitation energy of the 
spin-half {\it XXZ} model on the linear chain using the LSUB$m$ approximation
based on the $z$-aligned N\'{e}el model state, compared to results of
exact Bethe Ansatz and results of linear spin-wave theory.

\vspace{0.5cm}

\noindent
{\bf Figure 3:} CCM results for the ground-state energy of the spin-half 
{\it XXZ} model on the square lattice using the LSUB$m$ approximation
based on the $z$-aligned N\'{e}el model state, compared to quantum Monte 
Carlo (QMC) calculations \cite{qmc1}.

\vspace{0.5cm}

\noindent
{\bf Figure 4:} 
CCM results for the sublattice magnetisation of the spin-half 
{\it XXZ} model on the square lattice using the LSUB$m$ approximation
based on the $z$-aligned N\'{e}el model state.

\vspace{0.5cm}

\noindent
{\bf Figure 5:}
CCM results for the lowest-lying excitation energy of the spin-half 
{\it XXZ} model for the square lattice, using the LSUB$m$ approximation
based on the $z$-aligned N\'{e}el model state, compared to results of linear 
spin-wave theory.


\pagebreak
\pagestyle{empty}

\pagebreak

\pagebreak

\pagebreak

\pagebreak

\end{document}